# CAD Applications and Emerging Research Potential in Medical Imaging

Roshan P Mathews[1*]  and   Greeta Mathews[2**]

[*]Student Member IEEE.   [1]Electrical Engineering Department, IIT Palakkad, Kerala, India
[2]Asst Professor, Dept of Radiology, BR Ambedkar Medical College, Bangalore
[**]Correspondence Author- e-mail: drgreeta@gmail.com

*Abstract*— Computer Aided Detection (CAD) is a valuable technique for precisely interpreting medical images and it has a global business opportunity of about USD 1.8 billion. The current aspects with reference to the four sub stages such as image pre-processing, segmentation, feature extraction and classification and the future scope of CAD in medical imaging has been discussed in this paper. Many reviewers have emphasized the need for synergy between engineers and medical professionals for successful development of CAD systems and the current work is a move in that direction. The engineering aspects of the above four stages in four imaging modalities viz. computed tomography, magnetic resonance imaging, mammography and bone scintigraphy used in the diagnosis of five critical diseases have been discussed with a clinical background. Automatic classification of image can play an important role in preliminary screening of very critical ailments bringing down the cost of health care. Another recent advancement is using artificial intelligence and machine learning techniques. This paper reviews these engineering aspects with a view to explore the opportunities to researchers as well as the medical industry to offer affordable medical services with accessibility in even remote locations.

*Keywords— Computer Aided Detection (CAD), medical image processing, segmentation, feature extraction, classification*

## I. INTRODUCTION

The discipline of medical imaging has grown as an indispensible tool in identifying and recording the implicit conditions of the human body mostly in a non invasive manner. Computer aided detection (CAD) attempts to utilize the advancements in computational science to assist medical professionals in interpreting images leading to faster and effective diagnosis and are helpful in identifying anomalies that are sometimes difficult to notice [1 - 8]. Computer aided medical image analysis and identification of the abnormality involves a series of engineering activities e.g. image processing, pattern recognition etc. Different CAD systems handle different types of medical problems and varied types of data and images[9–17]. Hence any direct comparison based on a few parameters within the engineering domain may not be accurate. Joint project teams consisting of engineers and radiologists are the need of the hour to collaboratively develop systems that are useful in a clinical setup. In this study review of the following engineering aspects of CAD for a few common critical ailments has been done.

*a. Image pre-processing*: This step aims to improve the image quality by denoising, enhancing edges and other features.
*b. Segmentation*: In this step the image is partitioned to select the structures of interest.
*c. Feature extraction*: In the third step characteristic features of the segmented image is extracted for the next step.
*d. Classification*: Based on the extracted features the segmented objects are assigned to specific classes.

This paper is concerned about the engineering aspects in four imaging modalities viz. computed tomography (CT), magnetic resonance image (MRI), mammography and scintigraphy used in the diagnosis of lung cancer, intra cranial aneurysms, breast cancer, prostate cancer and bone metastasis. In addition to the review of the research, extensive discussions were held with domain experts, both engineers and radiologists, on the following broad research question:- what are the recent trends especially in engineering aspects of CAD systems and the computational techniques involved in the detection of a few of the most common cancers (lung, breast and prostate) and its bone metastasis and another critical pathology viz. intra cranial aneurysms. It is well known that in medical science, due to inherent heterogeneities in ailments and patients a uniform procedure can't be applied to all scenarios. An attempt has been made in this paper to comprehend current trends and future directions from a myriad techniques used in CAD systems. The review and discussions are based on the selected papers from Science Direct, PubMed, MEDLINE, NCBI data base, IEEE Xplore Digital Library etc. Recent studies were used for assessing the future trends in CAD.

## II. DIAGNOSTIC MODALITIES

Many research papers are available on CAD for various ailments such as cancer, vascular abnormalities like aneurysms (which if ruptures could be catastrophic), osteoporosis (weakened bone) with increased risk of fractures etc which can cause mortality or morbidity. Review papers in this area [2, 24, 59] indicate that the main focus of CAD research has been on lung [19], breast [20] and colon [21] cancer (though other organs like brain [22], liver [13], skeletal system [23], skin [9], cell features in hematological malignancies [10] etc. are also a subject of study in CAD research). As each of these ailments are separate medical specialties and beyond the scope of the current work, a few specific cases as shown in Table 1 are taken up and a brief description of the modalities is given below. Detailed discussions on the computer aided detection of these diseases will be taken up separately in section IV.

TABLE 1. DISEASES AND DIAGNOSTIC MODALITIES CONSIDERED

| Disease | Imaging Modality |
|---|---|
| Lung Cancer | CT |
| Intra Cranial Aneurysms | MRI |
| Breast Cancer | Mammogram |
| Prostate Cancer | MRI |
| Bone Metastasis | Bone Scan |

### A. Computed Tomography (CT)

CT scans very simply use an X-ray source that sends x-rays through the patient and the emerging beam that is attenuated by different structures differently is detected by detectors on the opposite side and the X- ray source and detector(array) move around the patient during the scan. The 4th generation scanners are rotate-stationary where the X-ray source moves around the patient with a stationary complete circle of detector array outer to the X-ray source. CT is an imaging modality that enables the clinician to non-invasively visualize the pathology within the patient. The CT scanner obtains images in the axial plane and the computer can use the data to generate sagittal and coronal and also 3 dimensional (3D) reconstructions. CT machines are generally designated by the number of slices it can obtain in one gantry rotation e.g 16 slice / 64 slice /128 slice CT etc, which means that the particular machine is capable of acquiring that many images per gantry rotation. The CT scanners of today, with an array of multiple detectors, also acquire data from a volume of the patient (spiral CT) rather than sequential axial images (this can also still be taken for specific indications) and this allows very short scan times, procedures like angiography to be done (long scan times would lead to loss of data due to contrast wash out from the vascular system) and information can be reconstructed in very thin slices.

*B. Magnetic Resonance Imaging (MRI)*

MRI is an imaging technique which is not based on X-rays or other ionizing radiations but uses magnetic field, radio frequency (RF) pulses and magnetic encoding gradients. This technology utilizes the spin of hydrogen ions (protons) present abundantly in the body. The strong external magnetic field produces longitudinal magnetization, pulses of radio waves cause transverse magnetization which precesses generating current which is received as a signal by the coil, and encoding gradients localize the signal within the patient. By varying the time interval between the RF pulses and the time after the pulse that the signals received are read, tissues will show different intensities (and therefore contrast in the image) depending on their longitudinal (T1) and transverse (T2) relaxation times.

*C. Mammography*

This technique uses X-rays of low-energy (to increase inherent contrast and enhance detail within the breast tissue and for better visualization of micro-calcifications some patterns of which raise high suspicion for malignancy) for screening of the breast and diagnosis of breast pathology. Screening mammography programs aim for the early detection of breast cancer as earlier detection improves prognosis. Patterns suggestive of malignancy on mammograms are a) mass with ill-defined or spiculated margins b) pleomorphic or linear branching microcalcifications in a cluster or with regional/ segmental distribution c) asymmetric density d) architectural distortion. The mammogram result is usually expressed as a BI-RADS (Breast Imaging Reporting And Data System) Category which ranges from 0 to 6, 0 denoting incomplete & needs further evaluation / comparison with previous studies, grade 1 denotes no abnormality detected, grade 2 is benign findings, for grade 3 follow up is needed, grade 4 is suspicious lesion needing FNAC (fine needle aspiration cytology)/ biopsy, grade 5 is highly suspicious for malignancy with suggested biopsy and 6 denotes biopsy proven malignancy.

A few other modalities can also be used for breast imaging. For example, ultrasound is particularly useful in young patients with dense breast tissue. It is also used for further assessment of a palpable abnormality not seen on mammogram. Characterization of a lesion seen on mammogram especially to determine its solid or cystic nature is yet another area where ultrasound can be used for breast imaging. Another modality is MRI, particularly suitable for young patients with 'BRCA mutations' (BRCA – BReast CAncer gene) who have dense breasts and higher chances of breast cancer in a younger age. MRI also finds use in suspected multicentric cancer and invasive lobular cancer or extensive Ductal Carcinoma In-situ (DCIS) (as the extent may be underestimated on ultrasound and mammograms and there is higher chance of bilateral disease). Ductography is used to rule out small intra ductal lesions like papillomas in cases presenting with nipple discharge and no abnormality detected with other modalities. As a recent trend, Positron Emission Mammography has been approved for assessment of breast.

*D. Scintigraphy*

A bone scan or bone scintigraphy is a bone imaging technique using nuclear medicine. It helps to diagnose pathologies like bone metastasis, fractures (subtle / stress fractures that may not be detected by traditional X-ray images), bone infection and show diffuse changes in metabolic bone disease. It uses small amounts of radioactive materials (most commonly technetium 99m phosphate labeled leucocytes) called radiotracers injected intravenously. The tracer tends to collect more where there is greater bone turnover (breakdown /repair). The radiotracer gives off gamma rays which are detected by a special gamma camera and the data is analyzed by the computer to generate images of the uptake. The areas of increased uptake are called hot spots and they show up on the scan as darker than other areas of bone.

## III. ENGINEERING ASPECTS OF DETECTION

The discussion on core engineering portion of CAD systems is taken up in this section with the help of a typical flow chart shown in Figure 1

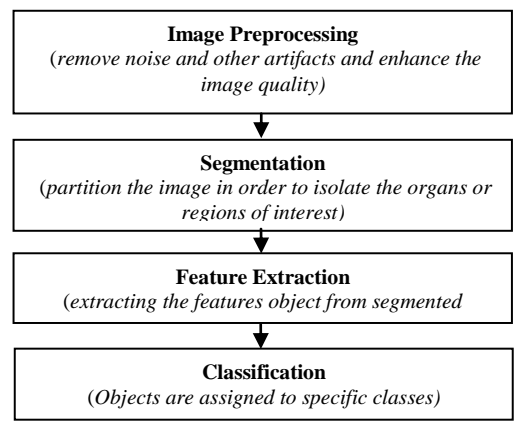

Fig. 1. Typical flow chart of a CAD System

In this scheme four major steps are involved in detection of ailments. The pre-processing stage involves cropping, noise removal etc. with a view to reduce the noise and improve the image for further processing [18, 25]. It is followed by image enhancement in terms of improvement of contrast, edges, separating objects from the background etc. Further segmentation is done to clearly identify the region of interest of study followed by feature extraction. The final stage of the process is classification. Though the above representation gives an overall idea of the sequence of activities, each of these stages employ various techniques in real life applications [26-28]. A brief discussion of each of these stages is given below:

*A. Image Pre processing*

Pre-processing in medical imaging generally refers to a set of image manipulations to improve its quality and hence ease of detection of ailments. This objective is achieved by applying suitable transformations or mathematical operations on the captured image [18,19, 29]. In the clinical setup, noise can mix with the image during its capture or transmission owing to a variety of reasons. Ensuring the best quality images is a basic necessity for developing CAD systems for medical application for assessing their detection capability. A brief discussion on image quality improvement techniques such as reduction of artifacts, denoising and image sharpening are taken up next.

*1) Image Artifacts Reduction*

Metallic implants appear as dark and bright streaks in CT and these artifacts spoil the quality of image. To improve the image, artifact reduction methods are available in latest CT machines. The algorithms for metal artifact reduction essentially recreate the original image in place of the corrupted ones using interpolation function from neighboring uncorrupted areas. Another way for metal artifact reduction is using dual-energy CT where data is captured at two different energy spectra thereby reducing the effects of beam hardening. Advantage with artifact reduction algorithms is that it can be done after image acquisition.

*2) Image Noise Reduction*

Image intensity disturbances of a random nature and coarse grains unintentionally get added during image capture or transmission. Impulse noise, additive noise, multiplicative noise etc are some common types of noise encountered in medical images. These noises misrepresent the true pixel values of the image and make it difficult to interpret the actual scenario intended for study. Hence denoising algorithms are applied to improve the image. Typically such algorithms smooth the image at overall level except the areas near boundaries of differential contrast.

*3) Image Sharpening*

The contrast at edges of features of interest plays a major role in image sharpness. It can be improved by using algorithms that increase edge contrast. There are many image processing techniques reported in the literature [52 - 55]. For example, sharpening filter is typically used in the mammogram in order to maximize the contrast between masses and the local background. Use of fuzzy histogram hyperbolization algorithm as a filter is reported to have provided better contrast[53].

From the literature, we can identify a few groups of preprocessing techniques. First one is denoising using different filters viz. mean filters, median filters [30, 31], wiener filtering [32], Gaussian filter [33], bilateral filtering [34] etc. Some researchers combined median filters with Laplacian filters [36, 37]. The second group is enhancing the edges of image structures. This includes various sharpening techniques including wavelet transform and others. Another trend is focusing on enhancing image contrast that includes techniques like histogram equalization. Another area is enhancement of 3D images. An example is the filters developed on the basis of eigen values of the Hessian matrix [38, 39]. Similar approach was reported by Frangi et al. [40] for enhancement of vessels in the image. The choice of preprocessing method is dependent on many factors including the quality of the image captured and requirement of the intended application. From the foregoing discussion we can infer that image processing techniques are very important in the development of CAD systems. The effectiveness of the subsequent stages and eventually the accuracy of detection is dependent on the quality of the images used. This is even true for manual applications where radiologists are reading the images and making the diagnosis.

*B. Segmentation*

Segmentation for different organs/structures in the image and matching with the anatomic data bank is the vital part of CAD. Basically in an automated process, algorithms are to be developed for identifying the edges or boundaries of the anatomical structures or organs. Unless the programme can recognize the objects of interest in an image, it is impossible to make an automated mining and retrieval of the required information from the image archives. One important challenge is decoding the image scans with missing edges or insufficient textural contrast between adjacent regions differentiating organs or body structures. Researchers [19,23,39] have studied such issues in segmentation and a few feasible techniques have been suggested. Compactness analysis, localization based on form, size, location with reference to near by structures or organs, identification based on the gray level with reference to the border of certain areas of interest etc. are some of the techniques used to segment medical images. In general, segmentation models could be viewed as deformable models and machine learning based models which classify according to certain classification approach [23, 24]. It can be further classified as supervised models or unsupervised models. Another approach is to use automatic seed selection and segmentation by region growing method.

In the context of Lung cancer detection research, Mahersia et al [19] gives a detailed account on the segmentation approaches. The CT image after the pre-processing stage is partitioned into different regions so that the pixels or voxels representing the lung tissues can be identified from the surrounding anatomical structures. 2 dimensional (2D) studies and 3 dimensional (3D) studies are becoming the main sections of image segmentation exercise [35, 39]. The 2D and 3D approaches as detailed by Mahersia et al [19] are shown in Figure 2 and Figure 3 respectively.

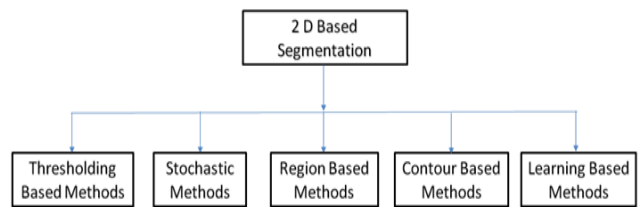

Fig 2  2D Segmentation method for Lung CT

The region of interest (ROI) denotes a specified area selected in the image for further quantitative analysis and extraction of the distinguishing features. The predefined areas in the image can be delineated based on certain criteria such as individual pixel value or average pixel value. The region of interest can be manually drawn to denote the geometrical shape of the organ of interest or it can be done in an automatic manner during segmentation by specifying the shape of the organ of interest or the anatomical structure we are concerned about.

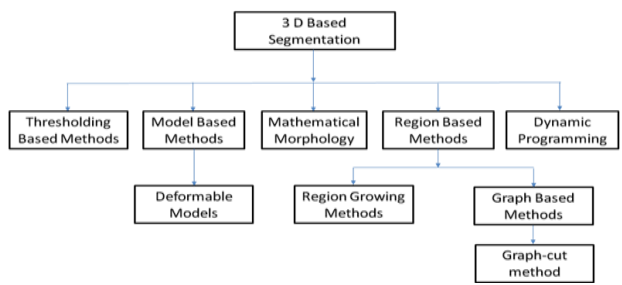

Fig 3  3D Based Segmentation methods

### C. Feature extraction, Classification & Clustering

The Features that are to be extracted from a segmented image are dependent on the case under investigation. Characteristic Features of lung nodule on CT is totally different from features of interest we are looking for in intra cranial aneurysms on brain MRI. Given the characteristic features of a case under investigation, the important element of the detection kernel is the classification algorithm that work based on the key features extracted from the regions of interest. There are classifier algorithms available for grey scale images and color images. Other properties of these images like texture can be another means of achieving classification. Location can be another basis of classification of images. For example, in lung cancer detection, lung nodules, if they exist, are to be extracted from the location of interest. For detection of anomalies leading to ailments like lung cancer, classification and clustering are the commonly used CAD technique. This is generally accomplished using techniques such as K-nearest neighbour, Fuzzy and neural network, linear discriminant analysis, support vector machines etc as shown in Figure 4

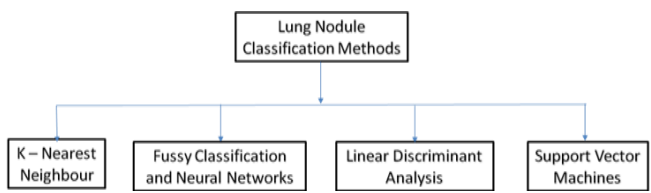

Fig 4  Important methods for Lung Nodule detection [19]

Another well known classification techniques is based on neural network, either in a supervised mode or unsupervised mode. Yet another promising approach is based on data mining schemes. Several related techniques employing statistical tools, visualization techniques, data and knowledge extraction frameworks etc can be clubbed with the data mining schemes. The artificial intelligence and machine learning techniques have tremendous potential in revolutionizing the CAD prospects in medical imaging field [25,37, 43-47].

## IV. CAD ANALYSIS OF DIFFERENT CONDITIONS

Five critical and most common ailments viz., lung cancer, intra cranial aneurysms, breast cancer, prostate cancer & metastatic bone cancer are selected for review and discussion on aspects of CAD systems. In addition, specific detectability issues of each ailment /modality is discussed from a clinical perspective.

### A. Lung cancer.

CAD with chest radiographs and CT scan is used for detection of lung nodules that may be either (a) malignant (primary or metastatic) or (b) benign. In a clinical set up, some nodules may be missed even on CT scans because of their location ie., within airways, adjacent to vessels, extreme periphery of lungs, small in size, faint or inhomogeneous internal architecture or being located adjacent to other lung abnormalities such as fibrosis or pneumonia etc. A typical CT image of lung nodule is shown in Figure 5.

CAD systems perform a very useful role in preliminary screening for detection of lung nodules in CT scans where a large number of studies used to be normal with only a few cases being positive. CT scans generate a large volume of images and computers being able to process large volumes of information in a small time would be able to assist the radiologist in reading of the images quicker by highlighting abnormalities and with the effect of a double reading of the study. During a study on detection of lung cancers, Armato et al. [41] reported that CAD system could detect 84 percent of the 38 missed cases. Also, when CT scan follow up of indeterminate lesions is done, computer analysis can reduce the time taken to compare the change, if any, in the nodules in terms of size, internal characteristics etc, particularly when the nodules are multiple. Computer can automatically tack the changes in size and complexity of a particular nodule in the corresponding CT sections.

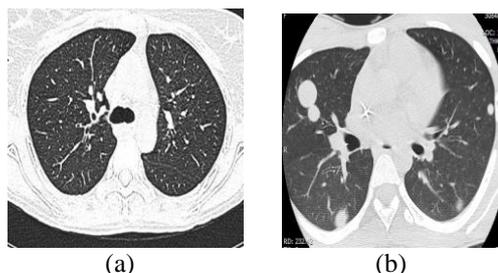

(a)  (b)
Fig 5 Typical CT images showing the difference between (a) Normal Lung (left) and (b) Lung Nodules

CAD systems are also used for differential diagnosis of lung nodules as benign and malignant eg. systems that integrate CT and positron emission tomography (PET) studies can help in characterization. Incorporating features such as margin characteristics, peak enhancement, internal architecture etc. into CAD techniques can help in characterization. With progress in related areas of technology, picture archiving and communication system (PACS) have become popular in clinical imaging service and data management which helps in easy retrieval of cases for comparison by CAD and also matching cases with similar lesions for machine learning.

*a. CAD Techniques for Lung Cancer:*

Lung cancer detection schemes using different approaches of image processing and machine learning have been studied by many researchers and a prominent model is suggested by Aggarwal et.al [42]. In the particular model, multiple parameters such as geometrical features, statistical parameters and the gray levels are extracted to decide the distinction between nodules and normal lung structure. This model uses optimal thresholding algorithm for segmentation and linear discriminant analysis for classification. Under this setup the model is reported to have achieved accuracy of 84% with sensitivity of the order of 97% and specificity 53%. Another CAD system model using convolution neural network as classifier to detect the lung cancer reported in 2016 by Jin et. al [43] obtained similar levels of accuracy (84.6%) but sensitivity was much less at 82.5%. But specificity of 86.7% was much better than that of the model suggested by Aggarwal et.al [42]. Higher detection accuracy of 90.7% was reported by another model [44] with median filter for pre processing stage and K-mean unsupervised learning algorithm for clustering. Further accurate model (accuracy = 94.12%) using fuzzy interference system trained with extracted features like area, mean, entropy, correlation, length of major axis and minor axis has been reported by Roy et.al [45]. This model is reported to have used image contrast enhancement with gray transformation also. Another model developed by Ignatious and Joseph [46] used Gabor filter in preprocessing and watershed segmentation. Using the neural fuzzy model and region growing method the model achieved a nodule detection capability of about 90% accuracy. However the models suffered from the serious limitation that they can not classify the detected lung nodule as a benign or a malignant nodule.

One of the major achievements in CAD system for lung cancer detection was the model proposed Gonzalez and Ponomaryvo [47] that could distinguish between benign and malignant nodules. This model uses Gabor filter in image preprocessing and marker controlled watershed method for segmentation. The region of interest is identified by the model using the priori information and the Hounsfield Unit (HU). It uses multiple characteristics such as area, perimeter, and eccentricity of the cancer nodules. For training the support vector machine in order to identify whether the given nodule is benign or malignant, a variety of characteristic features like area, circularity, eccentricity, fractal dimension etc. and key textural features such as mean, variance, skewness, energy, entropy, contrast etc. are used in the model. Even though this model [47] can classify cancer as benign or malignant, the need for prior information is a limitation. The model has been further refined by Suren Makaju et al [48] and the schematic view of the detection process using support vector machine is shown in Figure 6 below.

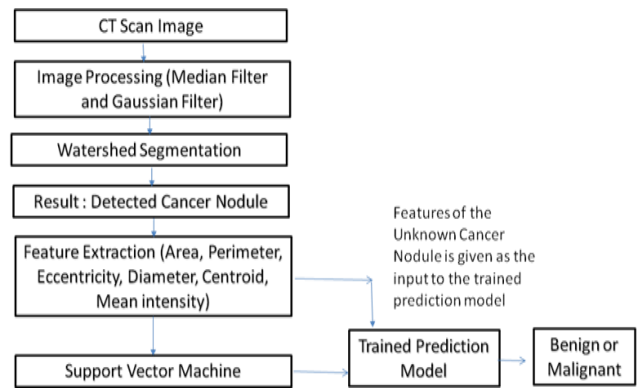

Fig 6 Lung Cancer Detection scheme [48]

*b. Specific aspects of CAD Detectability of Lung Nodules*

As discussed above, detectability being the main issue, a few aspects that affect the detectability of lung nodule are discussed below. Considering these factors would help to design robust CAD systems.

*(i) Size of lung nodules:* Size of the lung nodule is an important factor and the CAD systems are typically designed to detect lung nodules in a specific range of diameters. For example, in screening for lung cancer, the system may be designed to detect nodules greater than or equal to 5mm and upto 20mm as larger nodules are easily detected by the radiologist and smaller nodules may not be significant for malignancy in a screening setting. This approach ensures in reducing the false-positive rate. However for a patient with a known malignancy, even a nodule smaller than 5mm is also significant. Therefore, it is a point for improvement that the target range (say 5-20mm) should be adjusted according to the program and the flexibility to keep surveillance at smaller size nodules to be built within the CAD system.

*(ii) Non-solid or partly solid nodules:* Another issue for automatic detection is the type of nodule such as a solid nodule, a partly solid nodule, or as a non-solid (ground glass) nodule as it appears on CT. It has been observed that these type of non-solid or partly solid nodules may be early forms indications or precursors of a particular type of cancer called as adenocarcinoma. It is a possibility that these nodules may often go undetected while screening CT with CAD as most of the computer detection programmes are aimed at detecting solid nodules. Therefore, developing a robust CAD system for detection of non solid nodules is another important area of improvement.

*(iii) High false positives:* In a screening setting, only a small proportion of nodules will be malignant while most are benign. To overcome this high false positive rate for lung cancer (which has undesirable effects of repeated follow ups of patient or invasive procedures like biopsies for benign lesions), the CAD system may consider other methods such as nodule characterization (margins, enhancement peak, internal characteristics etc) as an auxiliary confirmation mechanism to be incorporated into the current schemes. Reasons for false positives may be bronchovascular structure, fibrotic scar, pleural thickening and other causes such as subtle respiratory/ cardiac motion artifacts.

*(iv) Co-existing lung disease:* Underlying disease like pneumonia, fibrosis may cause CAD systems to not register adjacent / contiguous nodules.

*B. CAD with MRA for Intra Cranial Aneurysms*

Rupture of intra cranial aneurysms leading to hemorrhage is a very critical situation for the patient and a noninvasive test called Magnetic Resonance Angiogram (MRA) can help in studying the blood vessels in the head and neck and pre-empt a very dangerous situation. In clinical practice, MRA often acts as a complementary procedure to MRI scan. Magnetic resonance angiography with computer aided detection (CAD) of intra cranial aneurysms reduces the time taken by radiologists in reading a study and improves the detection rates. A number of recent studies have shown that 3T MRI scanners were quite accurate for detection of even small (<5mm) intra cranial aneurysms. Figure 7 shows the typical images showing the normal Circle of Willis (left) and aneurysm.

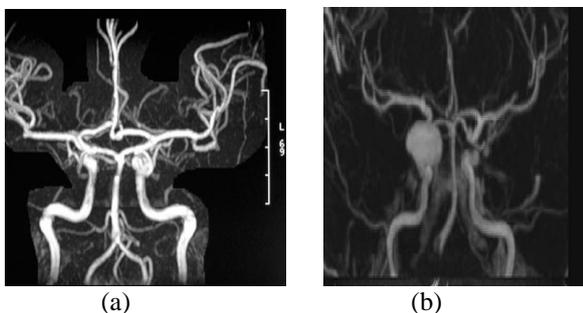

Fig 7 Typical MIP images showing the (a) normal Circle of Willis (left) and (b) aneurysm

MRA exhibits less sensitivity in the presence of overlapping vessels. The overlapping vessels often result in obscuring small aneurysms on maximum intensity projection (MIP) images and hence lead to failure of detection by radiologists. For achieving good sensitivity for detection it is recommended to assess both source and reconstructed images even at the cost of extra time spent. However, CAD systems can significantly reduce the time to read the study by highlighting abnormalities based on the parameters provided and a study by S. Miki et al [49] found that in many cases post review of CAD results, the radiologists had revised and updated their prior negative diagnosis to a positive confirmation. This was reported to have happened many times and the CAD diagnosis was found be correct to the extent of 92% and in the balance 8% case studies aneurysms found by the first radiologist with the help of CAD system had been negated after consultation with other specialists. This study [49] made it amply clear that CAD system was helpful for the detection of initially missed aneurysms than introducing false positives.

*a. CAD Methodologies for Intra Cranial Aneurysms*

MRA based computerized scheme for detection of intracranial aneurysms was developed by Arimura et.al [50]. At the image processing stage it uses a selective, multi-scale enhancement filter. The classification is based on linear discriminant analysis classifier. 207 cases of aneurysm images taken from three different types of MR equipment were used to evaluate the CAD model performance. The results were reported as having 96% sensitivity with 3.2 false positives per patient for detecting intracranial aneurysms in the range of 1 mm to 23 mm in size.

Another improved CAD system was reported by S. Miki et al [49] in 2016 with a provision to display top 3 likely lesion in the order of their likelihood. In another development Yang et al [51] presented a fully automated CAD that detects aneurysms on 3D time-of-flight MRA images. This system adopt automatic segmentation of intracranial arteries and locates the points of interest from the segmented vessels. It can also identify points of interest directly from the raw, (unsegmented) dataset. Regarding the performance of this CAD, it was reported that for a given operating point, overall sensitivity was 80% with three mean false positives per case. The sensitivity was 71% for aneurysms less than 5 mm size at the above test point. At another operating point the system achieved 95% overall sensitivity and sensitivity of 91% for aneurysms less than 5 mm, with nine mean false positives per case. The evaluation was based on 287 standard reference datasets which contained 147 aneurysms.

*b. Detectability aspects of intra cranial aneurysms*

From the practical point of view, the poor detection rates for cerebral aneurysms could be due to the following factors:

*(i) Size Effect:* Size is an important parameter affecting detectability of aneurysms and small (<5mm) or very small (<3 mm) aneurysms are difficult to get detected on maximum- intensity- projection images. There are reports that sensitivity drops to the level of 50% even for the experienced radiologists checking with standard datasets for detecting small aneurysms less than 5mm. Further, it was also reported that detectability is limited in the case of time-of-flight-MRA. The reason could be low signal intensity within the aneurysm due to slow flow. CAD has lot of potential for advancement in this area.

*(ii) Lack of 3T MR imaging systems*: Aneurysms can show variability in signal intensities on MR images. This is very likely when using a lower strength MR imaging system. The typical source of errors could be flow-related enhancement causing high signal on T1-weighted image, turbulence and extensive wall calcification causing signal void and blood clots of different ages causing heterogeneous signal intensities.

*(iii) Lack of optimal MR angiography imaging parameters*: Specialized post processing techniques, including volume rendering and a single-artery highlighting approach, may improve the detection of small cerebral aneurysms. Also, detection of aneurysms and reducing false positives is improved by analysing the images on a work station with ability to see the images from different views rather than static MIP images.

*(iv) Other Aspects*: In addition, small aneurysms may be missed on maximum intensity projection images, and false-positive diagnoses may occur as a result of vessel looping/ tortuosity, vessel overlap, atherosclerotic plaques, turbulent flow. According to studies, the site of the aneurysm was also a major factor that influenced its detection or lead to false positives.

## C. Mammography for Detection of Breast Cancer

The role of CAD is in improving the sensitivity by helping doctors to detect suspicious lesions as it decreases the likelihood of the radiologist making false negative diagnosis. Hence in clinical practice CAD performs the role of the second reader. The advent of CAD for clinical practice was in 1985 when it was introduced in the University of Chicago. Later on in 1998, after completing all the required trials and tests, United States Food and Drug Administration (FDA) approved the CAD system in Screening mammography. Subsequently many CAD systems came in to existence for clinical applications [68]. On mammograms, breast cancer may be seen as any of these conditions viz (a) mass, (b) microcalcification, (c) architectural distortion or (d) focal asymmetry. The computer detects these abnormalities and reports the results. The overall sensitivity of CAD was reported to be 91 % (139 of 152 cases) in the clinical study by Murakami et al [52]. In their study, CAD detected all the 47 cases (ie.,100 %) of cancers manifested as microcalcifications and all the 15 (ie.,100 %) that appeared as masses with microcalcifications. The detection rate of 98 % (62/63) was reported for non-calcified masses. Further, the study reported that 75 % (12/16) of the cases appeared as architectural distortions and 69 % (18/26) cases as focal asymmetry.

Cancers like invasive lobular cancers (ILC) and Ductal Carcinoma In-situ (DCIS) may be missed on mammography due to poor margins between the lesion and normal breast parenchyma. It was found that CAD detection wasn't influenced by the histopathology of cancer i.e the rates of detection was not poorer for these above described poorly marginated lesions. It is well known that detection of small lesions in screening is the key for early detection and management of breast cancer.

Murakami et al [52] showed that CAD with full field digital mammography improves early detection of tumours in the range of 1 to 10 mm size by identifying 83 % of cases (10 out of 12). For lesions larger than 10 mm size the sensitivity was 92 % (ie., 129 out of 140). Thus, CAD would help in reducing chances of missed cancers and thus improve the prognosis. CAD systems are also used with MRI and ABUS (automated breast ultrasound) for breast cancer detection and the results are reported to be excellent.

### a. Types of Tissues / Tumors in Breast Cancer:

In mammography, the term 'mass' means a space-occupying lesion commonly described in terms of its imaging features like margin, location, density and associated calcifications if any. The masses observed can be either benign or malignant. Normally, round shaped lesion with smooth, well defined margins / low-density are considered benign. On the contrary, high-density masses with spiculated or ill-defined margins are recognized as malignant. Other indications of underlying tumours are architectural distortion and focal asymmetry. A comprehensive account on the mammography images of circumscribed masses, spiculated masses and microcalcifications are reported in the work of Rafayah [53]. Typical images showing normal mammogram(left) and mass in breast(right) are shown in Figure 8 for illustration.

Microcalcifications are seen as tiny bright specks in the mammogram which are tiny calcium deposits. Their average size is about 0.3 mm according to the publication of Arianna et al [54] and they may present as an isolated finding or be associated with mass. Malignant microcalcifications are cluster of pleomorphic or linear branching type and focal / regional /segmental in distribution. If more than five micro calcifications are found in a cluster, that is suspicious for malignancy. Benign calcifications are macrocalcifications and appear larger in size (around 1–4 mm in size), coarse or popcorn type or peripheral calification with a lucent centre, similar in shape, rounded, diffusely scattered. Punctate, diffusely scattered, bilateral microcalcifications are usually benign.

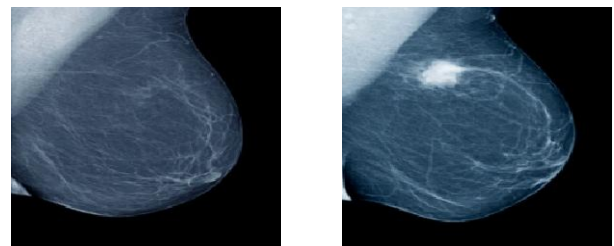

(a)           (b)
Fig 8 Typical images showing (a) Normal mammogram (left) and (b) mass in breast

### b. CAD Methodologies

After the acquisition of images, the mammogram images are pre-processed to eliminate or diminish the noise that may be present in the image. Filtering techniques can be utilized for this purpose. It also involves elimination of unwanted structures such as background. Further, to improve the quality of the image through increasing the contrast between structures, contrast enhancement techniques are adopted. Certain specific and novel pre-processing techniques have also been proposed in the literature for this purpose. The pre-processed image is further considered for processing. In certain cases, image transformation techniques are applied to transform the image to a different domain such that the region of interest (tumor in this case) is exposed more clearly. Wavelet transformation, curvelet transformation, Fourier transform etc. are commonly used for this purpose [55]. The process of image segmentation is then performed on the obtained transformed image or the pre-processed image or both to delineate the region of interest. The most commonly used segmentation technique is thresholding, where the region of interest is depicted with white pixels and the other regions with black pixels. Once the region is segmented, morphological operations are executed to remove some false positive regions and to clean the image. Pertaining to execution of image processing techniques alone, the process of detection ends here and the detected region of interest is highlighted as the tumor. In the case of incorporating pattern recognition techniques also, further steps have to be performed. To implement pattern recognition techniques, expressive features are extracted from the segmented region, pre-processed images, transformed images or all of them. Then, significant features are selected from the extracted features with respect to distinguishability. The selected features are then fed into machine learning procedures such as classification and clustering to generate a learning model that predicts if the image contains tumor or not. Classification techniques demand training data with ground truth for building the learning model while clustering techniques do not require ground truth. The pattern recognition procedures [20] can be adopted at pixel level, component level or image level. In the case of pixel level, the feature set represents the pixel values from various transformed domains whereas in the context of component level, the properties of the detected regions of interest such as its area, perimeter, mean etc., forms the feature set while in image level, aggregated features of the entire image form the feature vector. In pixel level analysis, the prediction refers to whether the pixel is a tumor pixel or background pixel; in component analysis, the prediction refers to whether the component is a tumor or false lesion whereas in image level the prediction denotes if the image is normal or diseased.

*c. Specific aspects of Detectability of Breast Cancer with mammography*

From the clinical perspective, detection rate of breast cancer by mammography may reduce due to the following factors:

*(i) Dense breasts*: the lesion would be obscured by increased density of normal breast parenchyma

*(ii) Architectural distortion, focal asymmetry/density* – As the edges are ill defined and contrast from the background density is poor, accurate measurement of size / extent is not possible. Hence, it is difficult for the radiologists or CAD system to detect these changes.

*(iii) Poor radiographic technique*: Following radiographic factors affect the accuracy of detection of anomalies in the mammogram. a] positioning b] inadequate compression c] immobilisation d] exposure factors

*(iv) Location*: Location of the tumor such as deep retroglandular /subareolar locations also affect the detectability.

*(v) Presence of Scars*: Malignancy in area where surgery or biopsy has been previously performed due to post procedure scarring (this should reduce with time on comparison studies)

*(vi) Small size*: As in other techniques, detectability in mammography is also affected by the nodule size. Very small nodules are unlikely to be picked up.

*D. CAD with MRI for Prostate Cancer*

As in the case of any other cancer, early detection and accurate diagnosis can improve survival chances in prostate cancer. Advent of multi-parametric magnetic resonance imaging (MP-MRI) has been a major advancement in prostate cancer imaging. This modality consists of T2-weighted sequences, dynamic contrast-enhanced MRI, diffusion-weighted sequences and spectroscopy. It is observed that large amount of data and differences in imaging sequences used affect detection due to the associated factors like as inter observer variability, underlying complexity and visibility of the lesions etc.. Properly designed and implemented CAD systems can improve quantitative assessment of the disease in a significant manner. Typical images showing the contrast between normal prostate MRI image and prostatic cancer are shown in Figure 9 for general understanding.

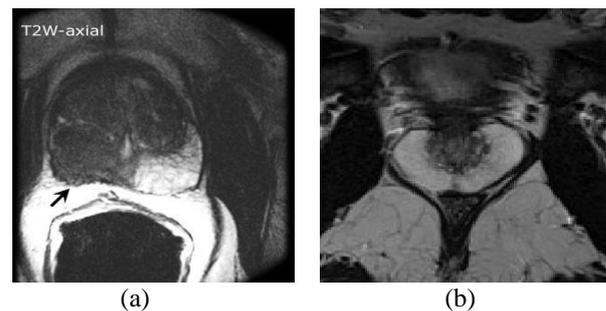

Fig 9. Contrast between (a) Normal prostate MRI image (left) and (b) prostatic cancer (right).

*a. MRI-based prostate CAD*

Dutch investigators have reported that MRI-based prostate cancer CAD system is performing well for detecting intermediate-to-aggressive prostate cancers noninvasively. Their study results from 130 prostate cancer patients reported sensitivity of more than 80% which is similar to the pickup by radiologists. According to the group from Radboud University Medical Center in Netherlands, when CAD was used as a second reader, the sensitivity almost reached 100% and the addition of CAD improved sensitivity for non aggressive versus aggressive tumors and particularly for PI-RADS 4 tumours [56]. Some other studies (Anika Thon et. al, [57]) were not able to replicate such a high success rate which indicates a need for further improvement.

*b. CAD Methodology*

In 2015, auntminnieeurope.com forum had reported the development of a functional prostate cancer CAD. Their feature extraction maintains quality and texture features and a voxel classifier is used for tumour candidate detection and segmentation. This feature, tested on 348 cases with cross validation, showed 82% sensitivity. CAD got validated using datasets and logistic regression analysis to arrive at a standalone CAD score, a radiologist score, and a combined score. 347 patients were used for training the system and 130 patients for the evaluation of CAD. For both groups, analysis was followed by MRI-guided biopsy and pathology. The results showed high sensitivity at a low false-positive rate. CAD performed at about the level of radiologists, and it provides complementary information so that combined, the result is significantly better.

*c. Specific aspects of detectability of Prostate Cancer*

Prostate cancer is hard to diagnose because the gold-standard diagnostic method is unreliable. With a high PSA (prostatic specific antigen) suspicious for malignancy, sometimes even transrectal ultrasound (TRUS) of prostate and biopsy may be negative. MRI can reduce the need for biopsy by 80%, but there are still cases where individuals with high PSA results undergo MRI, but the cancer cannot be found. To improve the chances of finding cancer, prostate cancer CAD was devised. A few years ago a functional prostate cancer CAD was developed in the Netherlands and showed good results. However, this CAD still needs fine tuning as the results could not be replicated by other authors and universal acceptability has not yet been achieved.

*E. Bone Scan of bony metastasis*

Detection and response of bony metastasis to treatment (radiotherapy or chemotherapy) is assessed by bone scans which have high sensitivity. Typical images showing normal bone scan and multiple sites of bony metastasis is shown in Figure 10.

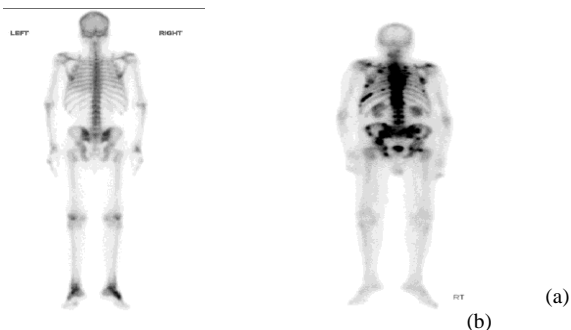

Figure 10 Typical images showing Normal (left) and Abnormal (multiple sites of bony metastasis) bone scan

Faster computational ability by machines reduces the time taken for comparing multiple areas of bone metastasis and to detect even subtle interval changes using a CAD scheme. Several studies have been reported in this area of research [69-74]. There are different CAD systems available for bone scan as given below:

**CADBOSS:** This is a software based system that comprises of basic parts of CAD algorithms for hotspot segmentation, feature extraction and/or selection and classification [58]. A level set active contour segmentation algorithm was used for the detection of hotspots. Moreover, a novel image gridding method was proposed for feature extraction of metastatic regions. An artificial neural network classifier was used to determine whether metastases were present. Performance evaluation of CADBOSS using an image database comprising of 30 non-metastases and 100 metastases revealed that the software system identified 120/130 images. The typical performance measures of the CADBOSS were reported as accuracy of 92.30% with sensitivity of 94% and specificity of 86.67%. The incorporation of CADBOSS system into the clinical framework was reported to have increased the clinician's success in detecting metastases from 95.38% to 96.9% [58] and help in decision making.

*(a) CAD using temporal subtraction technique*

In certain cases, CAD is used for detecting interval changes in successive whole-body bone scans employing temporal- subtraction methodology. It is obtained with a nonlinear image-warping technique [58]. This scheme consists of several steps viz., (a) initial image density normalization on each image, (b) image matching for paired images, (c) temporal subtraction by use of the nonlinear image-warping technique, (d) detection of interval changes by use of temporal-subtraction images, (e) image feature extraction corresponding to interval changes, (f) minimize false positives with the help decision criteria developed based on image extracted features and (g) display of the output for the highlighted interval changes. The temporal subtraction images obtained from previous and current bone scan and the correct detection of one cold lesion and two hot lesions by computer system as reported in [59] is shown in Figure 11 as an illustration of the capability and relevance of the CAD programme developed. This type of CAD software tools not only improves the detection rate but also make reporting much faster.

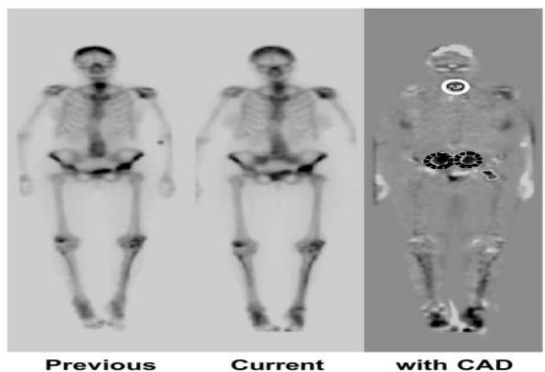

Fig 11 Illustration of temporal subtraction image [59].

## V. QUALITY OF DETECTION

The most important challenge in developing Computer Aided Detection (CAD) systems is ensuring that they are accurate and easily adaptable in clinical setup. In medical imaging, accurate diagnosis among other things depends mainly on image acquisition and its interpretation. The discipline of medical diagnosis has advanced multifold in the recent past, owing to evolutionary changes in many areas of digital imaging technology be it image quality, image capture systems, other digital systems. Recently data driven approaches and algorithms found a niche application in medical data in image interpretation. However a lot of things are yet to be realized in this area. For example, we found it very difficult to compare the performance of various CAD systems reported in literature as each system uses a different database and hence one to one comparison on a basis of reported accuracy and sensitivity won't represent the true performance and utility. To get a rough idea on the comparative evaluation of performance of various CAD systems, the papers by Petrick *et al* [60], Novak *et al*[61], and Kim *et al* [62] are useful.

Having touched upon the quality measures of some of the CAD systems along with the case studies discussed in the previous section, let us illustrate the real meaning of these performance measures such as sensitivity, specificity, accuracy etc using a typical scenario of tumor detection. The results obtained from CAD system typically fall into any one of the four scenarios namely, true positive (T+) denotes a tumor being correctly detected as tumor while true negative (T-) refers to a non-tumor being correctly detected as non-tumor. It may be noted that both (T+) and (T-) signifies a correct detection while false positive (F+) and false negative (F-) signifies a wrong detection. False positive (F+) refers to a non-tumor condition being predicted as tumor and false negative (F-) represents a tumor condition being denoted as non-tumor. With these notations, metrics such as accuracy, sensitivity, specificity, true positive rate etc., can be estimated as illustrated in Table II which is self explanatory.

Sensitivity of the CAD system is the ratio of number of cases where presence of pathology is correctly detected to the actual numbers of pathology present cases screened. The next performance measure is called specificity that gives an idea about the system's capability to rule out pathology when the patient is healthy. Accuracy is the overall measure defined as the ratio of sum of true diagnosis (both true positive and true negative) to the total number of cases diagnosed. From the point of view of the development of algorithms for diagnostic capability in CAD system, we can define a positive predictive value (PPV) as the ratio of the number of true positive cases to the total number of positive diagnosis made by the CAD system. Similarly, negative predictive value (NPV) is the ratio of the number of true negative cases to the total number of negative diagnosis made by the CAD system. Even though these two measures are not very significant in the clinical aspect, it is very much helpful for the engineers who develop and fine tune the algorithms.

TABLE II QUALITY PERFORMANCE MEASURES

| Patient's Clinical condition => | | Clinically ill (Tumor Present) | Clinically well (No Tumor) | CAD system Performance Measures |
|---|---|---|---|---|
| CAD System Result | Positive ($\alpha+\gamma$) | No. of true positive (T+) CAD report ($\alpha$) | No. of false positive (F+) CAD report ($\gamma$) | Positive Predictive Value PPV= $\alpha / (\alpha+\gamma)$ |
| | Negative ($\delta+\beta$) | No. of false negative (F-) CAD report ($\beta$) | No. of true negative (T-) CAD report ($\delta$) | Negative Predictive Value NPV= $\delta /(\beta+\delta)$ |
| Cumulative Clinical cases | | Actual No. of Tumor present cases = ($\alpha + \beta$) | Actual No. of Tumor Absent cases = ($\gamma + \delta$) | Total No of cases = ($\alpha + \beta + \gamma + \delta$) |
| CAD system Performance Measures | | Sensitivity = $\alpha / (\alpha + \beta)$ | Specificity= $\delta / (\delta + \gamma)$ | Accuracy = ($\alpha + \delta$) / ($\alpha + \beta + \gamma + \delta$) |

Processing time could be another quality characteristic of a CAD system which is the time taken by the system to give diagnostic results, that includes the processing time taken by sub modules such as image pre-processor, segmentation, features extraction, classification, evaluation etc. Ideally one should expect faster processing leading to shorter time elapsed per case without sacrificing the other quality characteristics discussed above.

While discussing the quality characteristics of a CAD system, one must keep a few background points in mind. Firstly the performance measures are liable to be a function of the test method. For example same CAD tested with different databases may show different performance levels. It is quite possible that a given detection algorithm is capable of producing 100% sensitivity result at 2 false- positives (F+) per image for a particular database. The same algorithm may report a sensitivity of 75% at two (F+) per image when applied to another database. Sometimes training dataset can also act as a confounding variable thus creating a large difference in tested performance level. This common problem can be rectified if features of the database images can be standardized in terms of its characteristics namely size of the lesion, contrast applied etc. Similarly a few subjective measures (eg. lesion subtlety rating, orientation of image etc.) can also serve as standardization parameter. The performance reported in the literature by various researchers may differ due to the non-standardized reporting format. One may report sensitivity in terms of percentage of detected abnormalities per image or the total number of abnormalities per case, which would result in different numerical values. Another source of variability in the reported performance of CAD system is use of different scoring criteria. Use of imprecise definitions, subjective classification rules etc make the comparison of CAD systems difficult. In addition, different investigators may use different detection criteria in evaluating CAD algorithms. Obviously detecting a malignant lesion is different from just reporting lesion/ abnormality in an image. Therefore, we must evaluate the computer programme in a holistic perspective and acknowledge that we need to compare it not only on the technical performance, but also the ease of integration into clinical practice.

VI. FUTURE TRENDS

CAD has gained considerable acceptance as a reliable tool that helps in the faster and more accurate detection of many diseases. CAD can help radiologists by highlighting subtle changes to avoid false negatives and quickly assess changes in serial follow up scans, thereby achieving accurate diagnosis [63]. The global CAD market size was roughly valued at USD 402.9 million in 2014 and its is expected to grow to the value of USD 3.9 billion by 2022 and to be worth USD 5.6 billion by the end of 2025. Many factors including increasing incidence of cancer, technological advances, rising awareness in the population for regular health check-ups due to the benefit of early detection of disease, technological advancements in CAD, its use in several diseases other than cancer and its easy integration with imaging technology are drivers for the growth of the global CAD industry. With newer techniques leading to large number of images requiring interpretation, CAD software tools support radiologists to read studies faster and more effectively. Applying CAD to newer areas in imaging, vigorous research and development, hospitals increasingly requiring automated workflow integration and PACS in their work environment to provide faster and more efficient care and increasing installation of CAD schemes in hospitals is further expected to boost this market. Some of the firms developing CAD solutions in the global market are Toshiba Medical Systems, Siemens Healthcare, Philips, GE Healthcare, Esaote, AGFA etc.

*A. Segment wise Trends:*

*a. CAD system to detect prostate cancer:*

As per the open source information available on the internet, the number of cancer patients is increasing globally as also prostate cancer cases among men. This increased the load on the healthcare system to have periodic screening which in turn sky rocketed the demand for CAD systems to be able to conduct periodic scans for a large population of men as part of the public health programme for prostate cancer detection.

*b. CAD system to detect breast cancer:*

According to the World Health Organisation data, breast cancer affects about 2.1 million women each year. Regular check-ups are necessary for early detection and management of breast cancer. CAD is expected to completely replace the role of the second reader radiologist in screening mammography. This would result in saving of the radiologists' time and also result in cost savings for the healthcare provider, insurer and therefore ultimately the patient. Currently, CAD in mammography is dominating the market due to the fact that breast cancer incidence is high and as there is a long period of research in this field with CAD software for mammography being the first commercially available computer detection programme. Recently many CAD systems were introduced for mammography and reported to have demonstrated good performance particularly in breast cancer detection for patients with dense breasts. CAD with other modalities (ultrasound, magnetic resonance imaging, positron emission mammography) are also picking up momentum and a lot of research work is going on at various institutes.

*c. CAD system to detect lung conditions.*

Lung cancer is among the top fatal diseases and has become a major cause of concern for public health authorities. Incidentally, mortality rate in women from smoking is increasing at a rapid rate. The role for CAD as second reader in areas such as low-dose lung cancer CT screening is promising with good performance seen with CAD for solid nodules. CAD research also seems promising in the detection and monitoring of other lung diseases such as pneumonia, acute pulmonary embolism, interstitial lung diseases etc. These areas are open for future research and development activities.

*d. CAD system with Other Modalities:*

Even though our discussion in this paper is limited to a few important imaging modalities, CAD finds application in other areas also as in the beginning of section II of this paper. To name a few X-ray, ultrasound (e.g.ABUS), electro cardiogram etc. use computer based detection. This list is expanding rapidly in covering a host of modalities that use CAD for fast and accurate detection of anomalies in various disciplines of medical science. One of the important advancements is in the detection of vertebral fractures due to osteoporosis on lateral chest radiograph.

B. *Future Trends in Technology*

Assessing future trends in computer application connected with medical imaging has been an interesting topic to engineers, medical equipment manufacturers, and clinicians as well. Kunio Doi [59] reviewed the prospects in medical imaging science and technology and published the significant progress made in the field of diagnostic imaging. He opines that the future of research in this area has a good potential and there is enough scope for improvements in improving image quality in various methodologies, developing interconnectivity between medical data bases, newer methods for data storage and development of algorithms for data mining and classification etc. leading to improved speed and reliability. To build a successful CAD program, he suggested close coordination between radiologists who interpret the images and engineers who improve the physics of image generation, detection and classification. Edwards *et al* [64] also emphasized this point regarding increased collaborations between medical experts and engineers to facilitate development for techniques and efficient algorithms. Hence the current work reported in this paper can be considered as a right move towards that direction.

Artificial intelligence and deep learning techniques [64-67] are likely to dominate the future research trends in CAD and probably transform it into the next higher domain called computer aided diagnosis (CADx). Further, Goncalves *et.al*, [1] conducted a systematic review of the evaluation of CAD systems and found that most studies explored breast and lung cancers. So they emphasize the need for CAD in diagnosis of other diseases that are responsible for human morbidity and mortality worldwide. Also most of the CAD systems are limited by the use of only one type of image. The need of the hour is to combine multiple images from various modalities to evaluate the patient's condition holistically. This may be an interesting area for future research.

Edwards *et al* [64] has reviewed the recent trends based on the research work reported in selected top journals and conferences. One of the major observations was the recent emergence of deep neural network and the associated machine learning algorithms used for computer-aided diagnosis (CADx) which is the next step after detection of anomalies in the image. In [64], 29 studies were reviewed to identify trends in this field. It was found that the research has focused mostly on cancer-related diseases while gynaecology / pediatrics have not received similar attention. Most of the current studies employ convolution neural network (CNN) architecture and it is expected to grow further in future.

Advancements in digital image acquisition and processing have helped health care business in general and particularly medical imaging industry to grow at a fast pace. The data generated in medical imaging is so large that the conventional data handling modalities may not be able to cope up. The state of the art big data approach and data mining techniques are essential to handle day to day requirements of medical image analysis. The opinion and judgment by human experts can be subjective and prone to error or likely to have variations from person to person. Automation is an effective method of alleviating subjectivity from the diagnosis process. However conventional automation techniques cannot handle the highly complex process of medical diagnosis through interpretation of images. High performance computers with self learning algorithms are required for extracting the relevant information and classifying it for making intelligent decisions leading to accurate, reliable and efficient diagnosis.

Deep learning has made a paradigm shift in computation such that it extracts the known features of the anomaly from the image as well as identify and construct new combination of features leading to easy and accurate diagnosis. This opens up an area of predictive diagnosis where based on the current condition of the patient revealed through the acquired images combined with other health indicators, the software can make an assessment of progression of disease and guide the doctors in effective patient care and management of the ailment. With the introduction of artificial intelligence (AI) and machine learning (ML) in medical imaging, engineers have a major role to play in medical image processing, image interpretation and classification, data fusion from various modalities, detection, diagnosis and progressive assessment of various ailments using different modalities.

Machine learning and artificial Intelligence techniques can also enhance the understanding of a given condition and interpret the impact of various contributing factors and implication of different treatment modalities. The research on the development and use of various algorithms and techniques for image processing, segmentation, classification, detection and diagnosis are growing at an exponential rate and all the aspects of their future trends and potential cannot be fully covered in this section. Interested readers are requested to peruse the list of relevant literature cited herein for further information. Notwithstanding the fact that there are many areas of uncertainties to be addressed and further refinements required to be made in many of the engineering aspects of detection and diagnosis, the recent developments in computational science and medical imaging have shown significant promise in automated detection and diagnosis. We are currently on the cusp of a new era which is just beginning in medical image diagnosis employing new knowledge in machine learning techniques initiated from the advances in deep learning.